\documentclass[prl,twocolumn,floatfix,showpacs,superscriptaddress]{revtex4}

\usepackage{amssymb}
\usepackage{amsmath}
\usepackage{amsfonts}
\usepackage{graphicx}

\newtheorem{theorem}{Theorem}

\newenvironment{proof}[1][Proof]{\noindent\textbf{#1.} }{\ \rule{0.5em}{0.5em}}

\begin{document}

\title{Adiabatic quantum computation in open systems}
\author{M. S. Sarandy}
\affiliation{Department of Chemistry, University of Toronto, Toronto, ON, 
M5S 3H6, Canada}
\affiliation{Instituto de F\'{\i}sica de S\~ao Carlos, 
Universidade de S\~ao Paulo, S\~ao Carlos, SP, 13560-970, Brazil} 
\author{D. A. Lidar}
\affiliation{Department of Chemistry, University of Toronto, Toronto, ON, 
M5S 3H6, Canada}
\affiliation{Department of Chemistry and Department of 
Electrical Engineering-Systems, University of Southern California, Los Angeles, CA 90089}

\begin{abstract}
We analyze the performance of adiabatic quantum computation (AQC)
under the effect of decoherence. To this end, we introduce an
inherently open-systems approach, based on a recent generalization of
the adiabatic approximation. In contrast to closed systems, we show
that a system may initially be in an adiabatic regime, but then
undergo a transition to a regime where adiabaticity breaks down. As a
consequence, the success of AQC depends sensitively on the competition
between various pertinent rates, giving rise to optimality criteria.
\end{abstract}

\maketitle

Adiabatic quantum computation (AQC) is a promising paradigm for quantum
information processing, which appears particularly well suited to physical
implementations \cite{Farhi:01}. In AQC, an algorithm is implemented via the
slow evolution of a time-dependent Hamiltonian $H(t)$. 
AQC schemes have recently been proposed
based on superconducting flux qubits~\cite{Kaminsky:04Grajcar:04}. An
experimental implementation of an adiabatic optimization algorithm using
nuclear magnetic resonance (NMR) techniques has already been reported
\cite{Stephen:03}.
Moreover, it has recently been shown that
AQC and the standard circuit model of quantum computation are
equivalent up to polynomial resource-overhead~\cite{Aharonov:04Kempe:04,Siu:04}.

The robustness of AQC against errors has recently been analyzed in several
contexts \cite{Childs:02,Shenvi:03RolandCerf:04Aberg:04}. An important consideration 
is that, if decoherence occurs
in the instantaneous eigenstate basis, then AQC can be intrinsically robust
against environmental noise provided one runs the algorithm at a temperature
that is small compared to the minimum gap~\cite{Childs:02}.
However, despite the importance of this result for the robustness of AQC,
the choice of the system eigenstate basis as a preferred basis may not
always be a good approximation (especially for non-Markovian environments), 
since it implicitly assumes that \emph{the environment keeps track of the 
Hamiltonian evolution}. Moreover, from a more general point of view, a 
methodology to systematically study AQC under
decoherence has not yet been developed. In this work, we introduce such a
methodology and analyze the performance of AQC under decoherence modeled by
a rather general class of master equations. Our approach is based on a
recently introduced adiabatic approximation genuinely conceived for open
quantum systems \cite{SarandyLidar:04}. We show that this framework 
can be used to provide the optimal run-time of adiabatic quantum algorithms.
This allows for the understanding of the performance and robustness of
open-system AQC. We illustrate our method by discussing the adiabatic
implementation of the Deutsch-Jozsa algorithm under dephasing.

\textit{Adiabaticity in open quantum systems}.--- Let us consider a quantum
system coupled to an environment, or bath, evolving under the
convolutionless master equation 
\begin{equation}
{\dot{\rho}}(t)=\mathcal{L}(t)\rho (t).
\label{eq:t-Lind1}
\end{equation}
An example of this class of master equations is given by (we use $\hbar =1$
units throughout) 
\begin{equation}
{\dot{\rho}} =-i\left[ H,\rho \right] +\frac{1}{2}%
\sum_{i}\left( [\Gamma _{i},\rho \Gamma _{i}^{\dagger
}]+[\Gamma _{i}\rho ,\Gamma _{i}^{\dagger }]\right) .
\label{eq:t-Lind2}
\end{equation}
Here $H(t)$ is the time-dependent effective Hamiltonian of the open system
and $\Gamma _{i}(t)$ are time-dependent operators describing the system-bath
interaction. Eq.~(\ref{eq:t-Lind2}) with time-\emph{in}dependent operators $%
\Gamma _{i}$ is usually referred to as the Markovian master equation, or
Lindblad equation \cite{Alicki:87}. 
In a slight abuse of nomenclature, we will refer to the time-dependent 
generator $\mathcal{L}(t)$ [Eq.~(\ref{eq:t-Lind1})] as the Lindblad super-operator 
and the $\Gamma _{i}(t)$ [Eq.~(\ref{eq:t-Lind2})] as Lindblad operators.

The key idea required to establish a natural adiabatic approximation for
open systems is to replace the concept of adiabatic evolution of eigenspaces
of the Hamiltonian by adiabatic evolution of the Jordan blocks of the
Lindblad super-operator~\cite{SarandyLidar:04}. 
In the super-operator formalism, the density matrix
for a quantum state in a $D$-dimensional Hilbert space
is represented by a $D^{2}$-dimensional
\textquotedblleft coherence vector\textquotedblright\ $|\rho \rangle \rangle
=\left( \rho _{1},\rho _{2},\cdots ,\rho _{D^{2}}\right) ^{t}$ and the
Lindblad super-operator $\mathcal{L}$ becomes a $D^{2}\times D^{2}$%
-dimensional supermatrix \cite{Alicki:87}. The master
equation~(\ref{eq:t-Lind1}) generates a non-unitary evolution, since
$\mathcal{L}(t)$ is non-Hermitian, and therefore, generally, non-diagonalizable. 
However, one can always transform $\mathcal{L}(t)$ into the Jordan canonical form,
where it has a block-diagonal structure.
This is achieved
via the similarity transformation $\mathcal{L}_{J}(t)=S^{-1}(t)\,\mathcal{L}
(t)\,S(t)$, where $\mathcal{L}_{J}(t)=\mathrm{diag}(J_{1},...,J_{m})$ is the
Jordan form of $\mathcal{L}(t)$, with $J_{\alpha }$ denoting the Jordan
blocks.
Instantaneous right $\{|%
\mathcal{D}_{\beta }^{(j)}(t)\rangle \rangle \}$ and left $\{\langle \langle 
\mathcal{E}_{\alpha }^{(i)}(t)|\}$ bases in the state space of linear
operators can always be systematically constructed such that they obey the
orthonormality condition $\langle \langle \mathcal{E}_{\alpha }^{(i)}(t)|%
\mathcal{D}_{\beta }^{(j)}(t)\rangle \rangle =\delta _{\alpha \beta }\delta
^{ij}$, and such that the Jordan block structure is preserved under the
action of the Lindblad super-operator, i.e., $\mathcal{L}(t)\,|\mathcal{D}%
_{\alpha }^{(j)}(t)\rangle \rangle =|\mathcal{D}_{\alpha }^{(j-1)}(t)\rangle
\rangle +\gamma _{\alpha }(t)\,|\mathcal{D}_{\alpha }^{(j)}(t)\rangle
\rangle $ and $\langle \langle \mathcal{E}_{\alpha }^{(i)}(t)|\,\mathcal{L}%
(t)=\langle \langle \mathcal{E}_{\alpha }^{(i+1)}(t)|+\langle \langle 
\mathcal{E}_{\alpha }^{(i)}(t)|\,\gamma _{\alpha }(t)$, with $|\mathcal{D}%
_{\alpha }^{(-1)}\rangle \rangle \equiv 0$ and $\langle \langle \mathcal{E}%
_{\alpha }^{(n_{\alpha })}|\equiv 0$ \cite{SarandyLidar:04}. Here subscripts enumerate Jordan
blocks ($\alpha \in \{1,...,m\}$), superscripts enumerate basis
states inside a given Jordan block ($i,j\in \{0,...,n_{\alpha }-1\}$, $%
n_{\alpha }$ is the dimension of the Jordan block), and $\{\gamma _{\alpha
}\}$ are the (generally complex-valued) Lindblad-Jordan (LJ)
eigenvalues.
Then, {\emph{an open quantum system is said to
undergo adiabatic dynamics when its Hilbert-Schmidt space can be decomposed
into decoupled LJ-eigenspaces with distinct, time-continuous,
and non-crossing instantaneous eigenvalues of $\mathcal{L}(t)$ }}\cite%
{SarandyLidar:04}. Just as in the closed-systems case, one can express the
condition for adiabaticity in terms of the total time of evolution. To this
end, we expand
$|\rho (t)\rangle \rangle =\sum_{\beta =1}^{m}\sum_{j=0}^{n_{\beta
}-1}p_{\beta }^{(j)}(t)\,e^{\int_{0}^{t}\gamma _{\beta }(t^{\prime
})dt^{\prime }}\,|\mathcal{D}_{\beta }^{(j)}(t)\rangle \rangle $. It is
convenient to express the variables in terms of the dimensionless time
$s=t/T$, where $T$ denotes the total evolution time. Then, adiabatic
dynamics in the interval $0\leq s\leq 1$ occurs if and only if the following
time condition is satisfied: $T\gg \max_{\alpha
}\{T_{\alpha }^{c}\}$, where $T_{\alpha }^{c}$ denotes the \emph{crossover
time} for the Jordan block $J_{\alpha }$~\cite{SarandyLidar:04}. 
For the particular case of one-dimensional blocks, which appears in
our example below, we have~\cite{SarandyLidar:04}  
\begin{eqnarray}
T_{\alpha }^{c} &=& \max_{0\leq s\leq 1} |\,\sum_{\beta \neq \alpha } [
Q_{\beta \alpha}(0)-Q_{\beta \alpha}(s)\,e^{T\,\Omega _{\beta \alpha}(s)}  \nonumber \\
&& +\int_{0}^{s}ds^{\prime }\,e^{T\,\Omega _{\beta \alpha
}(s^{\prime })}dQ_{\beta \alpha}(s^{\prime })/ds^{\prime } ] |,
\label{tcross}
\end{eqnarray}
where $\Omega _{\beta \alpha
}(s)=\int_{0}^{s}\omega _{\beta \alpha }(s^{\prime })\,ds^{\prime }$, $
\omega _{\beta \alpha }(s)=\gamma _{\beta }(s)-\gamma _{\alpha }(s)$ (the
gap between Jordan eigenvalues), $Q_{\beta \alpha}(s) \equiv V_{\beta \alpha
}(s)/\omega _{\beta \alpha }^{2}(s)$, and $V_{\beta \alpha }(s)=p_{\beta }(s)\,\langle \langle \mathcal{E_{\alpha }}(s)|
\frac{d\mathcal{L}(s)}{ds}|\mathcal{D_{\beta }}(s)\rangle \rangle$ (matrix elements of the
time-derivative of the Lindblad super-operator). Note that a quantity
analogous to $Q_{\beta \alpha}$ appears in the standard condition for adiabaticity in
closed systems \cite{SarandyLidar:04}. In the expression for $V_{\beta \alpha }(s)$, 
upper indices in $p_{\beta }^{(j)}(s)$ and in the basis vectors $\{|{\mathcal{D}_{\beta }^{(j)}(s)\rangle \rangle }\}$ 
and $\{\langle \langle \mathcal{E}_{\alpha }^{(i)}(t)|\}$ were removed because 
the Jordan blocks are one-dimensional.
The crossover time $T_{\alpha
}^{c}$ provides a decoupling timescale for each Jordan block: \emph{provided 
$T \gg T_{\alpha }^{c}$ the Jordan block $J_{\alpha }$ is adiabatically
decoupled from all other blocks associated to a different eigenvalue}. 

\textit{Performance of open-systems adiabatic quantum algorithms}.---
The performance of AQC under decoherence can be analyzed consistently within
the present picture of open-systems adiabaticity. In particular, the maximal
crossover time $\max_{\alpha }\{T_{\alpha }^{c}\}$ determined by Eq.~(\ref%
{tcross}) provides the time-scale over which the adiabatic approximation
holds. Provided the evolution is as slow as is set by this time-scale, the
density operator evolves separately in sets of Jordan blocks related to
distinct eigenvalues of $\mathcal{L}(t)$. Thus, if the initial density
matrix is associated to a certain set of instantaneous Jordan blocks, it
will remain associated to the same instantaneous set at all times. Note
that, if there is an overall growing exponential in the r.h.s. of Eq.~(\ref%
{tcross}), then the adiabatic behavior takes place over a finite time interval
and, afterwards, disappears. In this case, which is an exclusive feature of
open systems, \emph{we have the existence of a privileged time for adiabaticity}.
Having determined the adiabatic time interval, the performance of the
algorithm can be understood from the adiabatic density operator $\rho
_{a}(s,\lambda _{i},T)$, where $\lambda _{i}$ are the system-bath
coupling constants. This operator is obtained by solving the adiabatic master 
equation~(\ref{eq:t-Lind1}), where we disregard any coupling among Jordan blocks 
associated to distinct eigenvalues.

The final result, coming from $\rho_a(1,\lambda_i,T)$, will then 
depend on a competition between the adiabatic run-time $T$ and the
coupling constants $\lambda_i$. On the one
hand, the adiabatic approximation is favored for a certain time interval. On
the other hand, decohering processes tend to progressively destroy the
performance of the algorithm over time (intuitively, decoherence
causes broadening of the energy levels, until they overlap). This compromise between
adiabaticity and decoherence generates an optimal run-time for the
algorithm, which provides the optimal success probability for
given system-bath coupling strength. In agreement with this picture, \emph{an optimal
time has indeed been detected} in the experimental NMR AQC algorithm reported in
Ref.~\cite{Stephen:03}. 
Here, we provide a general 
explanation for such an optimal time in terms of the decoupling of the Jordan blocks of ${\cal L}(s)$. 

\textit{Constancy of the gap}.--- An important condition for the decoupling
of the Jordan blocks is the existence of gaps in the spectrum of LJ
eigenvalues $\{\gamma _{\alpha }\}$. This is relevant for AQC, where a
major concern is the scaling of the gap with problem input size. In fact, 
there have been indications that AQC may take an exponential time to solve certain 
hard instances of NP-complete problems due to vanishingly 
small gaps~\cite{vanDam:02}. A physical interpretation for the exponential delay has 
been proposed in terms of the quantum tunneling of a large spin during
the computation~\cite{Farhi:02}. In the closed-systems case, it is in principle possible \emph{to keep
the gap constant throughout the execution of AQC} via a ``unitary
interpolation'' scheme \cite{Siu:04}.
However, the tradeoff in using this method is that, in
general, it may require many-body interactions. 
In spite of this difficulty, schemes with a constant gap are an interesting 
possibility, since they constitute a favorable situation for closed systems 
AQC. Thus, it is natural to ask whether a constant gap setting may also be 
implemented within an open systems context. 
Our methodology for AQC can then be used to answer this question. 
In fact, we will see that this possibility persists in the open systems
setting only under very special conditions. We emphasize that the general
approach we introduced above applies to all interpolation schemes, 
in particular to standard, linear interpolation AQC \cite{Farhi:01,Childs:02}. 
The latter has the advantage of avoiding the many-body interactions associated 
with unitary interpolation~\cite{Aharonov:04Kempe:04,Siu:04}.

Let us first show that if the Hamiltonian changes by a unitary
transformation then the corresponding super-operator $\mathcal{H}(s)$ also
changes by a unitary transformation. The eigenvalues of $\mathcal{H}(s)$ are
given by the set of energy differences $\left\{ \epsilon
_{mn}(s)=E_{m}(s)-E_{n}(s)\right\} $ and the eigenvectors by the set $%
\left\{ |\psi _{m}(s)\rangle \langle \psi _{n}(s)|\right\} $. Therefore, if
the Hamiltonian giving rise to $\mathcal{H}(s)$ changes by a
unitary transformation $U(s)$, then the eigenvectors of $\mathcal{H}(s)$
evolve as $|\psi _{m}(s)\rangle \langle \psi _{n}(s)|=U^{\dagger }(s)|\psi
_{m}(0)\rangle \langle \psi _{n}(0)|U(s)$. Expressing them as vectors $|%
\rho_{mn}(s)\rangle \rangle$ in the state space of linear operators, 
we have that $|\rho_{mn}(s)\rangle \rangle =\mathcal{V}^{\dagger }(s)|%
\rho_{mn}(0)\rangle \rangle $, with $\mathcal{V}(s)\mathcal{V}%
^{\dagger }(s)=I$, which follows from the orthonormality of $|\rho_{mn}(s)\rangle \rangle$. 
Hence, from $\mathcal{H}(s)|\rho_{mn}(s)\rangle \rangle =\epsilon_{mn}(s)|\rho_{mn}(s)\rangle\rangle$, 
we obtain that $\mathcal{H}(s)=\mathcal{V}^{\dagger }(s)\mathcal{H}(0)\mathcal{V}(s)$. 
\begin{theorem}
\label{t1} Consider a Lindblad super-operator $\mathcal{L}(s)=\mathcal{H}(s)+%
\mathcal{R}(s)$, where $\mathcal{H}(s)$ [$\mathcal{R}(s)$] denotes the
Hamiltonian [decohering] component. If the Hamiltonian changes as
$\mathcal{H}(s)=\mathcal{V}^{\dagger }(s)\mathcal{H}(0)\mathcal{V}(s)$
with $\mathcal{V}$ unitary, then a sufficient condition for a constant spectrum of $%
\mathcal{L}(s)$ is $\mathcal{R}(s)=\mathcal{V}^{\dagger }(s)\mathcal{R}(0)%
\mathcal{V}(s)$. If the Jordan form of ${\cal L}(s)$ contains just one-dimensional Jordan blocks, 
this is also a necessary condition. In the case of time-independent $\mathcal{R}(s)$, this simplifies to 
$\left[ \mathcal{R},\mathcal{V}(s)\right] =0$ or
$\left[\mathcal{R},\mathcal{V}^{\dagger }(s)\right] =0$. Under these
conditions open systems AQC with unitary interpolation is possible.
\end{theorem}
\begin{proof}
Sufficiency: By assumption we have $\mathcal{L}(s)=\mathcal{V}%
^{\dagger }(s)\left[\mathcal{H}(0)+\mathcal{V}(s)\mathcal{R}(s)\mathcal{V}%
^{\dagger}(s)\right]\mathcal{V}(s)$. Therefore if $\mathcal{R}(0) = \mathcal{%
V}(s) \mathcal{R}(s) \mathcal{V}^{\dagger }(s)$ then $\mathcal{L}(s)=%
\mathcal{V}^{\dagger }(s)\mathcal{L}(0)\mathcal{V}(s)$. By inserting this
equation in the right-eigenvector equation $\mathcal{L}(s) |\mathcal{D}%
_{\alpha}(s)\rangle\rangle = \gamma_{\alpha}(s) |\mathcal{D}%
_{\alpha}(s)\rangle\rangle$, we obtain that the eigenvalues of $\mathcal{L}%
(s) $ are independent from $s$. The simplification in the case of time-independent ${\cal R}(s)$ is immediate. 
Necessity:  Assuming, in the eigenvector equation 
$\mathcal{L}(s) |\mathcal{D}_{\alpha}(s)\rangle\rangle = \gamma_{\alpha}(s) 
|\mathcal{D}_{\alpha}(s)\rangle\rangle$, that ${\cal L}(s)$ has a constant spectrum, we obtain 
$[\mathcal{H}(s)+\mathcal{R}(s)]|\mathcal{D}_{\alpha
}(s)\rangle\rangle=\gamma_{\alpha}(0)|\mathcal{D}_{\alpha}(s)\rangle\rangle$
$\Rightarrow$ $\mathcal{V}(s)\mathcal{R}(s)\mathcal{V}^{\dagger}%
(s)|\mathcal{{\tilde{D}}}_{\alpha}(s)\rangle\rangle=[\gamma_{\alpha
}(0)\mathcal{I}-\mathcal{H}(0)]|\mathcal{{\tilde{D}}}_{\alpha}(s)\rangle
\rangle$ where $|\mathcal{{\tilde{D}}}_{\alpha}(s)\rangle\rangle=\mathcal{V}%
(s)|\mathcal{D}_{\alpha}(s)\rangle\rangle$. But then, if the Jordan form of ${\cal L}(s)$ contains just 
one-dimensional Jordan blocks, 
we have that the set $\left\{|\mathcal{D}_{\alpha}(s)\rangle\rangle\right\}$ 
is complete and constitutes a basis in the state space of linear operators. Hence, it follows that 
$\mathcal{V}(s)\mathcal{R}(s)\mathcal{V}^{\dagger}(s)=\gamma_{\alpha}(0)\mathcal{I}-\mathcal{H}(0)=\mathcal{R}(0)$.
\end{proof}

Theorem~\ref{t1} implies that constant gaps in the spectrum of $\mathcal{L}(s)$ are {\it a priori}  
\emph{non-generic}. This places a limit on AQC with constant gap in the presence of decoherence.

\textit{Adiabatic implementation of the Deutsch-Jozsa (DJ) algorithm under
dephasing}.--- Given a binary function $f$ which is promised to be either
balanced or constant, the Deutsch problem is to determine which type the
function is~\cite{Deutsch:92}.
Here we construct an adiabatic implementation
for the optimized version of the algorithm~\cite{Collins:98}.
The input
state is
$|\psi (0)\rangle =|+_{1}\rangle \otimes \cdots \otimes
|+_{N}\rangle $, where $|\pm _{i}\rangle =(|0_{i}\rangle \pm |1_{i}\rangle )/%
\sqrt{2}$, with $\{|0_{i}\rangle ,|1_{i}\rangle \}$ being the computational
basis for the $i^{\text{th}}$ qubit (eigenstates of the Pauli matrix $\sigma
_{z}$).
The initial Hamiltonian is chosen such that its ground state is $|\psi (0)\rangle $, 
i.e., $H(0)=\omega\sum_{i=1}^{N}|-_{i}\rangle \langle -_{i}|$, where $\omega $ is the energy
scale. The Deutsch problem can be solved by a single computation of the function $f$
through the unitary transformation $U|x\rangle =(-1)^{f(x)}|x\rangle $ ($%
x\in \{0,1\}^{N}$)~\cite{Collins:98},
so that in the $\{|x\rangle \}$
(computational) basis $U$ is represented by the diagonal matrix 
$U=\mathrm{diag}[(-1)^{f(0)},...,(-1)^{f(2^{N}-1)}]$.
An adiabatic implementation requires a final Hamiltonian $H(1)$ such that
its ground state is $|\psi (1)\rangle =U|\psi (0)\rangle $. This
is accomplished by a unitary transformation on $H(0)$, i.e., $%
H(1)=UH(0)U^{\dagger }$ \cite{Siu:04}.
Then the final Hamiltonian encodes the solution of the Deutsch problem in its ground
state, which can be extracted via a measurement of the qubits in the
basis $\{|+\rangle ,|-\rangle \}$. 
A suitable interpolation between $H(0)$
and $H(1)$, which preserves the spectral gaps, can be defined by $H(s)={%
\tilde{U}}(s)H(0){\tilde{U}}^{\dagger }(s)$, where ${\tilde{U}}(s)=\exp
\left( i\frac{\pi }{2}sU\right)$.
The run-time of the closed-system version of the algorithm can be determined
from the standard adiabatic theorem, yielding
$T\gg \pi /2\omega $. This result is independent of $N$, as required.

We now analyze the effect of dephasing in the computational basis $\left\{
|0\rangle ,|1\rangle \right\} ^{\otimes N}$. For simplicity, we consider the
case of a single qubit, i.e., $N=1$. 
Dephasing is modeled by the Lindblad operator $\Gamma =\lambda \sqrt{\omega 
}\,\sigma _{z}$, where $\lambda $ is a dimensionless parameter denoting the
strength of the dephasing and the factor $\sqrt{\omega }$ is introduced to
make the energy scale explicit. Thus, expanding the coherence vector $|\rho
\rangle \rangle $ in the Pauli basis $\{I,\sigma _{x},\sigma _{y},\sigma
_{z}\}$, the Lindblad super-operator for the master equation~(\ref{eq:t-Lind2}) is found to be 
\begin{equation}
\mathcal{L}(s)=\omega \left( 
\begin{array}{cccc}
0 & 0 & 0 & 0 \\ 
0 & -2\,\lambda ^{2} & 0 & q(s) \\ 
0 & 0 & -2\lambda ^{2} & -r(s) \\ 
0 & -q(s) & r(s) & 0%
\end{array}%
\right) ,  \label{lsop}
\end{equation}%
where $r(s)=-\cos {\frac{\pi F}{2}s}$, $q(s)=\sin {\frac{\pi F}{2}s}$, with $%
F\equiv (-1)^{f(0)}-(-1)^{f(1)}$. 
In our DJ implementation the Hamiltonian super-operator evolves unitarily, i.e., 
$\mathcal{H}(s)=\mathcal{V}^{\dagger }(s)\mathcal{H}(0)\mathcal{V}(s)$. Explicit 
evaluation of $\mathcal{V}(s)$ yields that $\left[ 
\mathcal{R},\mathcal{V}(s)\right] =\left[ \mathcal{R},\mathcal{V}^{\dagger
  }(s)\right] =0$. Hence, it follows from Theorem~\ref{t1} that
(non-generically) the LJ spectral
gaps are constant in this example.
Interestingly, this property is not restricted to
dephasing in this example, but holds also, e.g., for spontaneous emission,
where $\Gamma \propto \sigma _{-}$. Indeed, the explicit evaluation of the
eigenvalues of $\mathcal{L}(s)$ shows that they are independent from $s$ and
given by $\gamma _{1}=0$, $\gamma _{2}=-2\omega \lambda ^{2}$, $\gamma
_{3}=\omega (-\lambda ^{2}-\sqrt{\lambda ^{4}-1})$, and $\gamma _{4}=\omega
(-\lambda ^{2}+\sqrt{\lambda ^{4}-1})$. These eigenvalues are non-degenerate
for $0<\lambda <1$ and define four one-dimensional Jordan blocks for the
Lindblad super-operator, denoted by $J_{\alpha }$ ($\alpha \in
\{1,...,4\}$)
(thus the condition in Theorem~\ref{t1} is both necessary and
sufficient).
Expanding the coherence vector as $%
|\rho (s)\rangle \rangle =\sum_{\beta =1}^{4}p_{\beta }(s)\,e^{T\gamma
_{\beta }s}\,|\mathcal{D}_{\beta }(s)\rangle \rangle $, where the $|\mathcal{%
D}_{\beta }(s)\rangle \rangle $ ($\gamma _{\beta }$) denote the eigenstates
(eigenvalues) of $\mathcal{L}(s)$, and using it in the master equation~(\ref%
{eq:t-Lind1}), we can show that the block $J_{1}$ is already decoupled from
the others. Therefore adiabaticity is related here to the decoupling of the
remaining three Jordan blocks.

\begin{figure}[ht]
\centering {\includegraphics[angle=0,scale=0.34]{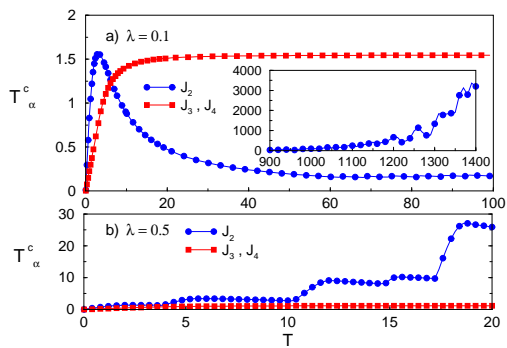}}
\caption{$T_{\protect\alpha}^c$ as a function of $T$ for the Jordan blocks 
$J_2$, $J_3$, and $J_4$. The inset of (a) shows that $J_2$ has non-negligible 
couplings for $T \sim 10^3$ (in units such that $\omega=1$).}
\label{f1}
\end{figure}

Next we compute the crossover times for decoupling of all the
Jordan blocks, so as to test for the adiabatic time interval, as defined by
the condition $T\gg \max_\alpha \{ T_{\alpha }^{c} \}$. We work in
units such that $\omega =1$. As anticipated above, one important result is
a finite time interval for adiabaticity. This is illustrated in Fig.~\ref{f1}, where we plot the
crossover time $T_{\alpha }^{c}$ as a function of the evolution time $T$ for
two values of $\lambda $. Observe that $T_{3}^{c},T_{4}^{c}$ asymptotically
approach a constant value, implying the decoupling of blocks $J_{3},J_{4}$
for sufficiently slow evolutions (large $T$) since the condition $T\gg
\max_\alpha \{ T_{\alpha }^{c} \}$ is satisfied. On the other hand, the
block $J_{2}$ can only decouple from the others during a finite interval 
[see inset of Fig.~\ref{f1}(a)].
While the adiabatic interval $T\gg T_{2}^{c}$ is large for $\lambda =0.1$,
it decays rapidly as the dephasing parameter $\lambda$ increases, 
as shown in Fig.~\ref{f1}(b).

In order to understand
the algorithm's performance we still need to analyze the adiabatic solution of the Lindblad
equation. Let us select $T$ such that, for given $\lambda$, adiabaticity is a
good approximation, i.e., we can disregard the Jordan block couplings. Then,
with $\rho (0)=(I+\sigma _{x})/2$, the straightforward solution of the
Lindblad equation yields $\rho (1)=[I+e^{-2\lambda
^{2}T}(-1)^{f(0)+f(1)}\sigma _{x}]/2$. The probabilities $p_{\pm }$ of
finding the system in one of the final states $\{|+\rangle ,|-\rangle \}$
are then $p_{\pm }=[1\pm e^{-2\lambda ^{2}T}(-1)^{f(0)+f(1)}]/2$. In the
closed-system case ($\lambda =0$) whether $f$ is constant or balanced is
determined, respectively, by $p_{+}=1$ or $p_{-}=1$. In the open-system
case, for each given value of $\lambda $, we can determine an optimal
run-time $T$, provided we impose a certain success probability for the
algorithm. For instance, take $\lambda =0.1$. Then, imposing a certainty of $%
90\%$ (either $p_{+}=0.9$ or $p_{-}=0.9$), we find $T\approx 11$. This
result is compatible with the adiabatic interval for $\lambda =0.1$, where
the condition $T\gg \max_\alpha \{ T_{\alpha }^{c} \}$ for $T=11$ is
relatively well satisfied, with $T_{2}^{c}\approx 0.82$ and $%
T_{3}^{c}=T_{4}^{c}\approx 1.43$. Therefore, for this dephasing scale, the
algorithm has a high probability of success. 
In order to generalize the results to many qubits, we consider 
$N$ independent dephasing operators $\Gamma_i=\lambda_i \sqrt{\omega} \sigma^z_i$ 
acting individually on each qubit. For $N=2$, it is easy to show that Theorem~\ref{t1} applies. 
We conjecture the validity of this result for any $N$ due to the simple diagonal form of ${\cal R}(s)$. 
A multi-qubit analysis can then be implemented similarly as done before. 
For $N$ qubits, the success probability may still be improved by repeating the algorithm execution several
times.

\textit{Conclusions}.--- We introduced and illustrated 
a general methodology to analyze the performance of AQC in open quantum systems
described by arbitrary convolutionless master equations.
We have shown
(Theorem~\ref{t1}) that a closed-systems unitary interpolation
(constant gap) scheme translates into an open system constant gap
scheme only under specific, non-generic assumptions.
The limited robustness of AQC in an
open-systems setting suggests that the development of quantum error
correction methods tailored to AQC is an important direction of future research.

\textit{Acknowledgments}.--- We gratefully acknowledge financial support
from CNPq-Brazil (to M.S.S.) and the Sloan Foundation (to
D.A.L.).

\end{document}